\documentclass[preprint,12pt]{aastex}

\usepackage[cp1251]{inputenc}
\usepackage[english]{babel}
\usepackage{amsmath}
\usepackage{amssymb}
\usepackage{natbib}
\usepackage[pdftex]{hyperref}

\hypersetup{
    a4paper,
    final,
    bookmarks=true,
    colorlinks=true,
    linkcolor=blue,
    citecolor=red,
    urlcolor=magenta,
    pdftitle={Electron-Positron Plasma Generation in a Magnetar Magnetosphere},
    pdfauthor={Denis Sobyanin},
    pdfsubject={Plasma Generation in a Strong Magnetic Field},
    pdfkeywords={Neutron Stars, Pulsars, Magnetars, Photon Splitting, Radio Emission}
}

\sloppypar
\hyperbaseurl{http://arxiv.org/abs/}
\slugcomment{\textwidth=2cm\it Astronomy Letters, 2007, Vol. 33, No. 10, pp. 660--672.$^*$}

\begin{document}
\title{\bf ELECTRON-POSITRON PLASMA GENERATION\\IN A MAGNETAR MAGNETOSPHERE}
\addcontentsline{toc}{section}{Title}
\author{\bf%
Ya. N. Istomin\altaffilmark{1,2\dag} and D. N. Sobyanin\altaffilmark{1\ddag}}

\affil{
$^1${\it Moscow Institute of Physics and Technology,\\Institutskii per. 9, Dolgoprudnyi, Moscow oblast, 141700 Russia}}
\affil{
$^2${\it Lebedev Institute of Physics, Russian Academy of Sciences,\\Leninskii pr. 53, Moscow, 119991 Russia}}
\vspace{2mm}
\hspace{50mm}{Received April 11, 2007}
\sloppypar
\vspace{2mm}

We consider the electron-positron plasma generation processes in the magnetospheres of magnetars --- neutron stars with strong surface magnetic fields, $B_0\simeq10^{14}{-}10^{15}$~G. We show that the photon splitting in a magnetic field, which is effective at large field strengths, does not lead to the suppression of plasma multiplication, but manifests itself in a high polarization of $\gamma$-ray photons. A high magnetic field strength does not give rise to the second generation of particles produced by synchrotron photons. However, the density of the first-generation particles produced by curvature photons in the magnetospheres of magnetars can exceed the density of the same particles in the magnetospheres of ordinary radio pulsars. The plasma generation inefficiency can be attributed only to slow magnetar rotation, which causes the energy range of the produced particles to narrow. We have found a boundary in the~$P{-}{\dot P}$~diagram that defines the plasma generation threshold in a magnetar magnetosphere.



\vfill
\noindent\rule{8cm}{1pt}\\
{$^\dag$ E-mail: $<$\email{istomin@lpi.ru}$>$}\\
{$^\ddag$ E-mail: $<$\email{sobyanin@gmail.com}$>$}
\\{\it $^*$Translated from Pis'ma v Astronomicheskii Zhurnal, 2007, Vol. 33, No. 10, pp. 740--753.}
\clearpage
\section*{INTRODUCTION}
\addcontentsline{toc}{section}{Introduction}

Magnetars are neutron stars with superstrong surface magnetic fields, $B_0\simeq10^{14}{-}10^{15}$~G. This is two or three orders of magnitude higher than the field strength typical of radio pulsars, $10^{12}$~G. Such magnetic fields for magnetars follow from the standard estimate of $B_0\simeq2(P{\dot P}_{-15})^{1/2}\,10^{12}$~G that is used for radio pulsars and that is based on the fact that the neutron star rotation energy is spent on the generation of magnetodipole radiation, which owes its origin to the magnetic field rotation. Just as radio pulsars, magnetars spin down, ${\dot P}>0$, but with a much higher rate, ${\dot P}\simeq 10^{-10}{-}10^{-12}$, while the typical spindown rate for radio pulsars is ${\dot P}\simeq 10^{-15}, {\dot P}_{-15}\simeq 1$. The rotation periods~$P$ of magnetars are also longer than those of radio pulsars and lie within the range $P\simeq 5{-}10$~s. The application of the magnetodipole formula is more justified for magnetars than for radio pulsars, whose magnetospheres must be filled with a dense electron–positron plasma that shields the magnetodipole radiation \citep{BeskinEtal1993}. In addition, the observation of an absorption line in the X-ray spectrum of the magnetar SGR~1806--20 \citep{IbrahimEtal2002}, which is interpreted as a proton cyclotron resonance near the stellar surface, yields the same estimate of the magnetic field strength for magnetars.

The term ``magnetar'' was introduced by \citet{DuncanThompson1992} for soft gamma repeaters (SGRs). At present, two groups of objects are classified as magnetars: SGRs and anomalous X-ray pulsars (AXPs), which currently number 5 and 9, respectively. SGRs yield gamma-ray bursts with energies up to $3.5\times10^{46}$~erg, which was observed on December~27, 2004, in SGR~1806--20; AXPs also exhibit bursts but in the X-ray energy range. SGRs and AXPs have similar periods and magnetic fields, but ${\dot P}$ for SGRs is slightly larger, ${\dot P}\simeq 10^{-10}$. Therefore, SGRs may be considered to be younger objects, with ages of $\simeq 10^3$~yr, than AXPs, whose ages are $\simeq10^{4}{-}10^{5}$~years.

It is important to note that the magnetic fields of magnetars also determine their energy activity. Indeed, the energy stored in the magnetic field of a magnetar,
\begin{equation*}
W_B=\int \frac{B^2}{8\pi}dV=10^{45}\left(\frac{B_0}{10^{14}\hbox{ G}}\right)^2\left(\frac{R}{10\hbox{ km}}
\right)^3\hbox{ erg}
\end{equation*}
exceeds the neutron star rotation energy,
\begin{equation*}
W_R=2\pi^2IP^{-2}=2\times10^{44}\left(\frac{I}{10^{45}\hbox{ g cm}^2}\right)\left(\frac{P}{10\hbox{ s}}
\right)^{-2}\hbox{ erg}.
\end{equation*}

This is not the case for radio pulsars: $W_R\gg W_B$. In addition, the X-ray luminosity of AXPs, $L_x\simeq 10^{35}{-}10^{36}\hbox{ erg s}^{-1}$, is much higher than the rotation energy being lost, $E_R=4\pi^2IP^{-3}{\dot P}\simeq 10^{32}{-}10^{33}\hbox{ erg s}^{-1}$. This means that the energy being released is not the neutron star rotation energy, as is the case for radio pulsars, but the magnetic field energy stored in the star. How the magnetic field energy of magnetars is released and where such strong fields in them originate from still remains a puzzle. In addition, it is unclear why the spindown of the star is still observed if its rotation is not the energy source. Strong magnetic fields can be generated by magnetic dynamo mechanisms \citep{DuncanThompson1992,BonannoEtal2006} and can result from the penetration of the radiation field into the neutron star crust with a high but finite conductivity \citep{Istomin2005}.

The absence of radio emission from magnetars in the standard range of frequencies at which radio pulsars are observed, $\nu\simeq 0.5{-}3$~GHz, suggests that there is no dense plasma in the magnetospheres of magnetars whose flows would generate radio emission, as is the case in the magnetospheres of radio pulsars. However, low-frequency ($60{-}110$~MHz) radio observations at the Pushchino Radio Astronomy Observatory revealed emission from some magnetars \citep{MalofeevEtal2004,MalofeevEtal2005}, although its characteristics differ from those of radio pulsars.

The goal of this paper is to investigate the generation of electrons and positrons in the magnetospheres of neutron stars with superstrong magnetic fields which are present in magnetars. We wish to ascertain why no dense plasma is generated in magnetars and what suppresses its effective generation.

For magnetars, the following critical magnetic field strength is a characteristic one:
    \begin{equation}
    \label{criticalField}
        B_c=\frac{m^2c^3}{e\hbar}\approx4.414\times10^{13}\text{ G},
    \end{equation}
where $m$~is the electron mass, $e$~is the electron charge, $c$~is the speed of light, and $\hbar$~is the Planck constant. When the magnetic field strength becomes comparable to~\eqref{criticalField}, the quantum nature of physical processes begins to manifest itself. If we consider the motion of
an electron in such a field, then only its momentum component parallel to the magnetic field direction remains continuous. The transverse momentum component is quantized in such a way that the expression for the total particle energy is
    \begin{equation}
    \label{quantumEnergy}
        \varepsilon_n^2=1+p^2+2nB,
    \end{equation}
where $p$ is the longitudinal particle momentum component and $n=0,1,\ldots$ is the number of the Landau level occupied by the particle. In what follows, we will measure the magnetic field strength in units of~$B_c$~\eqref{criticalField} and take the electron mass $m$, the Compton wavelength of the electron $^-\!\!\!\!\lambda=\hbar/mc\approx3.86\times10^{-11}$~cm, and its ratio to the speed of light $^-\!\!\!\!\lambda/c\approx1.29\times10^{-21}$~s as the units of mass, length, and time, respectively. Formally, this means that~$\hbar={^-\!\!\!\!\lambda}=c=1$. Thus, the length, velocity, time, energy, and wave vector will be measured in units of $^-\!\!\!\!\lambda$, $c$, $^-\!\!\!\!\lambda/c$, $mc^2$, and $1/^-\!\!\!\!\lambda$, respectively.

Two distinctly different regions are known to be identified in the magnetospheres of rotating neutron stars: the regions of open and closed magnetic field lines. In the region of closed field lines, the particles rotate synchronously with the field because of high plasma conductivity and magnetic field freezing-in. However, along the open field lines, the particles can move freely and escape from the neutron star magnetosphere.
Continuous plasma outflow from the magnetosphere requires the presence of electron–positron pair generation processes compensating for it.

In this paper, we consider the generation of relativistic electrons and positrons by photons in the superstrong magnetic fields of magnetars, the photon splitting in such fields, and its influence on the pair generation. By taking this into account, we derive expressions for the particle distribution function and density and find a criterion for effective plasma generation.
\section*{KINETIC EQUATION FOR RELATIVISTIC PARTICLES}
\addcontentsline{toc}{section}{Kinetic Equation for Relativistic Particles}

The electron–positron plasma generation processes in the magnetosphere of a neutron star are described by the kinetic equation
    \begin{equation}
    \label{generalKineticEquation}
        \frac{\partial F}{\partial t}+\textrm{div}(F{\bf v})+\frac{\partial(F{\dot {\bf p}}{\bf b})}{\partial p}=Q+S.
    \end{equation}
Here, $F$ is the particle distribution function in longitudinal momentum~$p$ or in longitudinal Lorentz factor~$\gamma$, which in dimensionless units coincides with the energy~$\varepsilon_n$~\eqref{quantumEnergy} at~$n=0$, ${\bf v}$~is the particle velocity, ${\bf b}={\bf B}/B$~is a unit vector along the magnetic field, $Q(F)$~is the operator that describes the generation of particles by high-energy photons, and $S(F)$~is the operator
that describes the scattering of particles through the emission of photons by them.

A particle with a Lorentz factor~$\gamma$ moving along a field line with a radius of curvature~$\rho$ at a given point emits photons called curvature ones. The emission probability $P_\pm^{(c)}(\gamma,k)$ of a curvature photon with energy~$k$ per unit time is given by the expression \citep{SokolovTernov1983}
    \begin{equation}
    \label{curvatureEmissionProbability}
        P_\pm^{(c)}(\gamma,k)=\frac{1}{2\sqrt{3}\,\pi}\frac{\alpha}{\gamma^2}\,\varphi_\pm\Big(\frac{k}{k_c}\Big),
    \end{equation}
where $\alpha=e^2/\hbar c=1/137$ is the fine-structure constant,~$P_+^{(c)}(\gamma,k)$ is the emission probability of a $\parallel$-polarized photon whose electric field vector lies in the plane passing through the vectors~${\bf k}$ and~${\bf B}$, and~$P_-^{(c)}(\gamma,k)$ is the emission probability of a $\perp$-polarized photon whose electric field vector is orthogonal to this plane. The characteristic energy~$k_c$ of the emitted curvature photons is
    \begin{equation}
    \label{curvatureEnergy}
        k_c=\frac{3}{2}\frac{\gamma^3}{\rho}.
    \end{equation}
The functions~$\varphi_\pm(x)$ are defined as follows:
    \begin{equation*}
    \label{phiPlusMinus}
        \varphi_\pm(x)=\int\limits_x^\infty K_{5/3}(y)\,dy\pm K_{2/3}(x),
    \end{equation*}
where $K_\nu(x)$ is the Macdonald function. Using asymptotic representations of the Macdonald function, we can easily derive asymptotic expressions for
the functions~$\varphi_+(x)$ and~$\varphi_-(x)$ that are useful in the subsequent analysis:
    \begin{align}
        \varphi_+(x)&=
        \begin{cases}
            2^{-1/3}3\,\Gamma\!\!\left(\!\displaystyle\frac{2}{3}\!\right)x^{-2/3}\quad\text{for }x\rightarrow0\\            \sqrt{\displaystyle\frac{2\pi}{x}}\,e^{-x}\quad\text{for }x\rightarrow\infty;
        \end{cases}    \label{phiPlusAsymptotics}\\
        \varphi_-(x)&=
        \begin{cases}
            2^{-1/3}\Gamma\!\!\left(\!\displaystyle\frac{2}{3}\!\right)x^{-2/3}\quad\text{for }x\rightarrow0\\
            \displaystyle\frac{\sqrt{{2\pi}}}{3}\,x^{-3/2}\,e^{-x}\quad\text{for }x\rightarrow\infty.
        \end{cases}    \label{phiMinusAsymptotics}
    \end{align}

Note that Eq.~\eqref{curvatureEmissionProbability} is valid if the classical theory of synchrotron radiation is applicable. The criterion for its applicability is
    \begin{equation}
    \label{classicalSyncrotronTheoryApplicabilityCriterion}
        \frac{\gamma^2}{\rho}\ll1.
    \end{equation}
Since the Lorentz factor of the particles~$\gamma$ and the radius of curvature of the field lines~$\rho$ in the magnetospheres of neutron stars are typically~$\gamma\lesssim10^7$ and~$\rho\gtrsim10^{19}$, in dimensionless units (note that in these units, $1000\text{ km}\approx0.26\times10^{19}$), condition~\eqref{classicalSyncrotronTheoryApplicabilityCriterion} is always satisfied. The satisfaction of criterion~\eqref{classicalSyncrotronTheoryApplicabilityCriterion} means that the characteristic energy of the emitted photons accounts for only a small fraction of the initial particle energy (see~\eqref{curvatureEnergy}).

\citet{Klepikov1954} was the first to show that a single-photon generation of electron–positron pairs is possible in a magnetic field. Let a photon with energy $k$ move at an angle $\chi$ to the direction of magnetic field ${\bf B}$. The photon absorption coefficient due to electron–positron pair creation in the case of a weak magnetic field, $B\ll1$, for two directions of photon polarization is then \citep{TsaiErber1974}
    \begin{equation}
    \label{weakFieldPhotonAttenuationCoefficient}
        \kappa_{\parallel,\perp}(B,k,\chi)=b_{\parallel,\perp}\alpha B\sin\chi\exp\Big(-\frac{8}{3}\frac{1}{Bk\sin\chi}\Big),
    \end{equation}
where
    \begin{equation*}
        b_\parallel=\frac{\sqrt{3}}{4\sqrt{2}}\approx0.31,\qquad        b_\perp=\frac{\sqrt{3}}{8\sqrt{2}}\approx0.15.
    \end{equation*}
Equation \eqref{weakFieldPhotonAttenuationCoefficient} is applicable when $Bk\sin\chi\ll1$. We should also take into account the fact that the pair
creation is possible only when a photon with energy $k$ propagating along the tangent to the magnetic field line at the point of emission reaches some threshold angle $\chi_t$ at which the transverse component of the photon wave vector is equal to 2,
    \begin{equation}
    \label{thresholdAngle}
        k\sin\chi_t=2.
    \end{equation}

If, alternatively, we consider the case of a strong magnetic field, $B\gtrsim1$, characteristic of magnetars, then we should take into account the fact that the electron–positron pairs are generated in such a way
that either both particles of the produced pair are at the zeroth Landau level if the pair-generating photon is $\parallel$-polarized or one particle of the pair is at the zeroth Landau level and the other particle is at the first Landau level if the pair-generating photon is $\perp$-polarized \citep{SemionovaLeahy2001}. In the former case, the photon absorption coefficient is
    \begin{equation}
    \label{strongFieldParallelPolarizationPhotonAttenuationCoefficient}
        \kappa_\parallel(B,k,\chi)=\frac{2\alpha B}{k^2\sin\chi}\left(\frac{k^2\sin^2\chi}{4}-1\right)^{-1/2}\exp\Big(-\frac{k^2\sin^2\!\chi}{2B}\Big)
    \end{equation}
and the threshold angle $\chi_t$ meets to the same condition~\eqref{thresholdAngle}.
In the latter case, the absorption coefficient~$\kappa$ and the angle~$\chi_t$ are
    \begin{equation}
    \label{strongFieldPerpendicularPolarizationPhotonAttenuationCoefficient}
        \kappa_\perp(B,k,\chi)=\frac{\alpha B}{p_{01}k^2\sin\chi}\left(1+p_{01}^2+\varepsilon_0\varepsilon_1\right)\exp\Big(-\frac{k^2\sin^2\!\chi}{2B}\Big),
    \end{equation}
    \begin{equation*}
    \label{strongFieldPerpendicularPolarizationThresholdAngle}
        k\sin\chi_t=1+\sqrt{1+2B},
    \end{equation*}
where
    \begin{equation*}
    \label{p01}
        p_{01}=\left(\frac{k^2\sin^2\chi}{4}-1-B+\frac{B^2}{k^2\sin^2\!\chi}\right)^{1/2},
    \end{equation*}
    \begin{equation*}
    \label{E0AndE1}
        \varepsilon_0=\left(\frac{k^2\sin^2\chi}{4}-B+\frac{B^2}{k^2\sin^2\!\chi}\right)^{1/2},
        \quad
        \varepsilon_1=\left(\frac{k^2\sin^2\chi}{4}+B+\frac{B^2}{k^2\sin^2\!\chi}\right)^{1/2}.
    \end{equation*}

Let us now find the form of operator $Q$ in the
kinetic equation~\eqref{generalKineticEquation}, which describes the generation of electron–positron pairs. Let $q_0(k)$ be the number
of photons emitted per unit volume per unit time. The opening angle of the emission cone is inversely proportional to the particle Lorentz factor~$\gamma$. Since the energy of the particles is fairly high, all of the emission may be assumed to be directed along their motion. The distribution function $f_0(k)$ of the emitted photons will then be proportional to the radius of curvature $\rho_0$ of the magnetic field line at the point of emission:
    \begin{equation}
    \label{initialPhotonDistributionFunction}
        f_0(k)=\rho_0q_0(k).
    \end{equation}
The function $f_0(k)$ is normalized in such a way that $f_0(k)\,dk$ gives the number of photons with an energy in interval $dk$ about $k$. Obviously, if there is no photon absorption, then the distribution function is conserved along the direction of emission. If, alternatively, the electron–positron pair generation takes place, then the photon distribution function $f(k)$ at the point under consideration is smaller and can be expressed in terms of the photon distribution function $f_0(k)$ at the point of emission and the absorption coefficient $\kappa$ as
    \begin{equation}
    \label{photonDistributionFunction}
        f(k)=f_0(k)\exp\Bigl(-\int\limits_{l_t}^l\kappa\bigl(B(l'),k,\chi(l')\bigr)\,dl'\Bigr),
    \end{equation}
where $l$ is the distance between the point under consideration and the point of emission and $l_t$~is the so-called threshold distance at which the electron–positron pair generation begins.

Assuming that the magnetic field in the neutron star magnetosphere in the polar region of interest near the surface is a dipole one, let us introduce the cylindrical $(r,z)$ coordinates in such a way that $r$ denotes the distance from the dipole axis to the point of observation and~$z$ denotes the distance from the dipole center coincident with the stellar center to a
given point along the dipole axis. Expanding the well-known expression for the shape of a dipole magnetic field line~$r_p=R\sin^2\theta_p$ specified in polar $(r_p,\theta_p)$ coordinates at small polar angles $\theta_p$, we find that
    \begin{equation}
    \label{fieldLineForm}
        r=\frac{1}{\sqrt{R}}z^{3/2},
    \end{equation}
where $R\gtrsim R_c$ is the distance from the stellar center at which a given field line will cross the $z=0$~plane when extended as a dipole one, $R_c=c/\Omega$~is the light-cylinder radius, and $\Omega$~is the pulsar rotation rate. We derive an expression for the radius of curvature of the
field line from Eq.~\eqref{fieldLineForm}:
    \begin{equation}
    \label{curvatureRadius}
        \rho=\frac{4}{3}\sqrt{zR}.
    \end{equation}
Assuming that $z=R_s$ and $R=R_c$, where $R_s\approx10$~km is the neutron star radius and $R_c\approx3\times10^5$~km is the characteristic light-cylinder radius for a pulsar with a period $P=6$~s, we obtain the radius of curvature $\rho\gtrsim10^3$~km mentioned above. Note that
Eqs.~\eqref{fieldLineForm} and \eqref{curvatureRadius} are valid to terms $O(z/R_c)$, which may be disregarded with a sufficient accuracy in the conditions under consideration.

Using Eqs.~\eqref{fieldLineForm} and \eqref{curvatureRadius}, we can easily obtain the dependence of the angle~$\chi$ between the directions of magnetic field~${\bf B}$ and photon wave vector~${\bf k}$ on the
distance~$l$ traversed by the photon:
    \begin{equation}
    \label{chiOnLGeneralDependence}
        \chi(l)=\frac{l}{\rho_0}\frac{1}{1+l/z_0},
    \end{equation}
where $\rho_0$ is the radius of curvature of the magnetic field line at the point of emission and $z_0$ is the distance to this point from the dipole center. Since we used the fact that the angle between the directions of the magnetic field and the photon wave vector is small, we can replace $\sin\chi$ by $\chi$. Below, we will assume that $l\ll z_0$. This means that
    \begin{equation}
    \label{chiCriterion}
        \chi(l)\ll\chi_{\max}=\frac{z_0}{\rho_0},
    \end{equation}
where $\chi_{\max}$ is the maximum possible angle between the photon wave vector and the magnetic field direction. When criterion~\eqref{chiCriterion} is met, we may ignore the dependence of the magnetic field strength~$B$ on the distance traversed by the photon when integrating in Eq.~\eqref{photonDistributionFunction} and assume (see~\eqref{chiOnLGeneralDependence}) that
    \begin{equation}
    \label{chiOnLDependence}
        \chi(l)=\frac{l}{\rho_0}.
    \end{equation}

The number of electron–positron pairs produced per unit time per unit volume at the point of absorption is
    \begin{equation}
    \label{pairProductionRate}
        \frac{dN}{dt}=\int\limits_{k_t}^\infty f(k)\kappa(B,k,\chi)\,dk,
    \end{equation}
where, as in the case of Eq.~\eqref{thresholdAngle}, the threshold energy
of a photon capable of producing electron–positron pairs, $k_t=2/\chi$, is defined.

According to \citet{Beskin1982}, the photon energy for a weak field in the case of electron–positron pair generation is distributed almost evenly between the electron and the positron. It thus follows that the energy of the produced particle is uniquely related to the angle between the photon wave vector and the magnetic field direction:
    \begin{equation}
    \label{gammaOnChiDependence}
        \gamma=\frac{1}{\chi},
    \end{equation}
where $\gamma$ is the Lorentz factor of the particle after it passes to the zeroth Landau level. For a strong field, $B\gg1$, as we will see below, the electron–positron pair generation threshold matches~\eqref{thresholdAngle}, since all photons will be $\parallel$-polarized due to the photon splitting in a strong magnetic field when this threshold is reached. In addition, the pair generation in this case will take place at the threshold point at angles~$(\chi-\chi_t)/\chi_t\ll1$. Therefore, the magnitude of the pair particle momentum in the reference frame where~${\bf k}\perp{\bf B}$ is close to zero. Hence, using the inverse Lorenz transformation, we can easily find that equality~\eqref{gammaOnChiDependence} is also valid for~$B\gg1$.

Thus, we find from \eqref{initialPhotonDistributionFunction}, \eqref{photonDistributionFunction}, \eqref{pairProductionRate}, and \eqref{gammaOnChiDependence} that the operator~$Q$ describing the particle source in the kinetic equation~\eqref{generalKineticEquation} is
    \begin{equation}
    \label{Q}
        Q=\rho_0\!\int d\chi\,\delta\Big(\gamma-\frac{1}{\chi}\Big)\int\limits_{2/\chi}^\infty dk\,q_0(k)\,\kappa(B,k,\chi)\exp(-\rho_0\!\int\limits_{2/k}^\chi\kappa(B,k,\chi')\, d\chi'),
    \end{equation}
where $B$ is the magnetic field strength at the point of absorption and the absorption coefficient~$\kappa$ is defined by Eq.~\eqref{strongFieldParallelPolarizationPhotonAttenuationCoefficient} for the strong magnetic fields of magnetars or by Eq.~\eqref{weakFieldPhotonAttenuationCoefficient} for the weak magnetic fields of ordinary pulsars. In Eq.~\eqref{photonDistributionFunction}, we passed from integration over the distance~$l'$ to integration over the angle~$\chi'$ using~\eqref{chiOnLDependence}.

\citet{GurevichIstomin1985} gave the form of the operator~$S$ describing the scattering of particles through the emission of curvature photons by them:
    \begin{equation}
    \label{S}
        S=\frac{2\,\alpha}{3\rho^2}\frac{\partial}{\partial\gamma}\left[\gamma^4F
        +\frac{55}{32\sqrt{3}\,\rho}\frac{\partial(\gamma^7F)}{\partial\gamma}\right].
    \end{equation}
Equation~\eqref{S} is valid when criterion~\eqref{classicalSyncrotronTheoryApplicabilityCriterion} is met. These authors pointed out that the operator~$S$ gives only a small contribution compared to the operator~$Q$. Therefore, it may be disregarded when the electron–positron pair generation processes are considered.

The last term on the left-hand side of Eq.~\eqref{generalKineticEquation} is responsible for the particle acceleration. We will assume that this acceleration takes place in a small region above the neutron star surface where the so-called inner gap is formed \citep{RudermanSutherland1975}. A significant accelerating electric field exists in this region, but outside it this field is shielded by the generated plasma and no particle acceleration takes place. It thus follows that we may disregard the term under consideration and take into account the particle acceleration as a boundary condition at the neutron star surface for the particle distribution function,
    \begin{equation}
    \label{boundaryCondition}
        F(\gamma)=N_0\delta(\gamma-\gamma_0),
    \end{equation}
where $N_0$ is the electron–positron plasma density at the neutron star surface and~$\gamma_0\simeq10^7$ is the Lorentz factor that the particles acquire as a result of their acceleration in the inner gap.

As regards the outer gap \citep{ChengRuderman1977}, if it emerges in the magnetosphere of a magnetar, then the electron–positron plasma generation
processes in it do not differ from those in an ordinary radio pulsar, since it is located far from the surface where the magnetic field is weak

Thus, the kinetic equation~\eqref{generalKineticEquation} in the case under
consideration transforms to
    \begin{equation}
    \label{KineticEquation}
        \frac{\partial F}{\partial t}+\textrm{div}(F{\bf v})=Q,
    \end{equation}
where the operator $Q$ is defined by Eq.~\eqref{Q}.
\section*{PHOTON SPLITTING IN A STRONG MAGNETIC FIELD}
\addcontentsline{toc}{section}{Photon Splitting in a Strong Magnetic Field}

The photon splitting processes become important in the strong magnetic field of a magnetar. It should be noted that the splitting probabilities are very sensitive to the photon polarization. Based on the CP invariance and the energy–momentum conservation law with vacuum polarization, \citet{Adler1971} showed that $\perp\rightarrow\parallel\parallel$~decay is crucial in moderately strong fields, $B\ll1$, and at low photon energies. Before the appearance of the paper by \citet{Usov2002}, the case of $\parallel\rightarrow\perp\parallel$ and $\perp\rightarrow\perp\perp$ decays in the calculations of magnetar magnetospheres, in general, was not ruled out \citep{HardingEtal1997,BaringHarding2001}. However, \citet{Usov2002} showed that Adler’s selection rules are valid in fields of arbitrary strengths. Thus, the decay of $\parallel$-polarized photons below the electron–positron pair generation threshold~\eqref{thresholdAngle} is strictly forbidden. Hence, it would be incorrect to consider a polarization-averaged photon distribution function and to use a polarization-averaged attenuation coefficient in a strong field.

Let $f_\parallel(k)$ and $f_\perp(k)$ be the photon distribution functions in energy~$k$ for two directions of polarization. We assume that the splitting of only $\perp$-polarized photons is possible. However, the photon need not decay into two photons with equal energies. We will describe the splitting probability of a $\perp$-polarized photon with energy $k_1+k_2$ into two $\parallel$-polarized photons with energies $k_1$ and $k_2$ per unit time by a function~$w(k_1,k_2)$.

The system of equations that describes the evolution of the distribution functions for photons of two polarizations is
    \begin{equation}
    \label{photonSplittingSystem}
        \begin{aligned}
            &\dot{f}_\perp(k)\!\!\!\!&=&-\frac{1}{2}f_\perp(k)\int\limits_0^k w(k-k',k')\,dk',\\
            &\dot{f}_\parallel(k)\!\!\!\!&=&\int\limits_0^\infty w(k,k')f_\perp(k+k')\,dk'.
        \end{aligned}
    \end{equation}
The first equation in \eqref{photonSplittingSystem} describes the splitting
of a $\perp$-polarized photon with energy~$k$ into two $\parallel$-polarized photons with energies~$k'$ and $k-k'$. The second equation describes the generation of a $\parallel$-polarized photon with energy~$k$ through the decay of a photon with a higher energy~$k+k'$ and the detachment of a photon with energy~$k'$. We immediately note that time differentiation for the photon distribution functions is equivalent to distance differentiation, since we assume that~$c=1$. In this case, the splitting probabilities per unit time and per unit length of the distance traversed by the photon are equal.

It follows from the first equation in~\eqref{photonSplittingSystem} that the
total photon splitting probability per unit time is
    \begin{equation}
    \label{photonSplittingTotalProbability}
        W(k)=\frac{1}{2}\int\limits_0^k w(k-k',k')\,dk'.
    \end{equation}

Let a photon with a wave vector~${\bf k}$ be emitted by relativistic particles propagating along the magnetic field lines at some point in the direction of their motion. \citet{BaringHarding1997} provide an approximation of the total photon splitting probability per unit length in the limit~$B\gg1$ as a function of the photon energy~$k$ with an accuracy of~$\sim1.5\%$:
    \begin{equation}
    \label{photonSplittingTotalProbabilityApproximation}
        W(k,\zeta)=\frac{2\alpha^3}{\pi^2}\frac{\zeta^6}{k}\left(\frac{4}{271}-\frac{1}{124}\zeta
        +\frac{9}{247}\zeta^2-\frac{14}{297}\zeta^3+\frac{1}{3461}\exp(5\zeta)\right).
    \end{equation}
Here, we introduced a variable~$\zeta=k\chi/2$, where~$\chi$ is the angle between the vectors ${\bf k}$ and ${\bf B}$  related to the distance traversed by the photon by Eq.~\eqref{chiOnLDependence}. The equation that describes the change in the distribution function of $\perp$-polarized photons is
    \begin{equation}
    \label{perpendicularPhotonDecay}
            f_\perp(k,\zeta)=f_\perp(k,0)\exp\Bigl(-\frac{2\rho_0}{k}\int\limits_0^\zeta W(k,\zeta')\,d\zeta'\Bigr),
    \end{equation}
where $f_\perp(k,0)$ is the distribution function of $\perp$-polarized
photons at the point of emission. As the photon moves, the variable~$\zeta$ increases and becomes equal to unity when the threshold of electron–positron pair generation by $\parallel$-polarized photons \eqref{thresholdAngle} is reached. Directly integrating Eq.~\eqref{photonSplittingTotalProbabilityApproximation} and substituting~$\zeta=1$ yields
    \begin{equation}
    \label{thresholdValueOfPerpendicularPhotonDistribution}
            f_\perp(k,1)=f_\perp(k,0)\exp\Bigl(-3.57\times10^{-3}\frac{4\alpha^3}{\pi^2}
            \frac{\rho_0}{k^2}\Bigr).
    \end{equation}

Thus, we may assume that all $\perp$-polarized photons had decayed before the threshold of pair generation by $\parallel$-polarized photons was reached, when the the expression in the exponent $\gtrsim1$. Therefore, the inequality $0.56\times10^{-9}\rho_0/k^2\gtrsim1$ must hold. Substituting the characteristic radius of curvature of the magnetic field lines, $\rho\gtrsim10^{19}$, into this inequality yields
    \begin{equation}
    \label{thresholdValueOfKForPhotonSplitting}
        k\lesssim10^5.
    \end{equation}

Note that inequality~\eqref{thresholdValueOfKForPhotonSplitting} always holds for typical conditions in the magnetosphere of a magnetar, since the maximum energy of the curvature photons has a characteristic value of~$k\simeq10^4$ for a particle Lorentz factor~$\gamma\simeq10^7$ (see~\eqref{curvatureEnergy}). Above this energy, the curvature-photon distribution function decreases exponentially (see asymptotics~\eqref{phiPlusAsymptotics} and \eqref{phiMinusAsymptotics}).

We see that as a result of the splitting of $\perp$-polarized photons for magnetars, all photons with be $\parallel$-polarized when the threshold angle~\eqref{thresholdAngle} is reached.

\citet{BaierEtal1996} showed that the total photon splitting probability increases as the electron–positron generation threshold~$\zeta=1$ is approached. In this case, the differential photon splitting probability near the threshold is almost constant, i.e., the photon can decay into any two lower-energy photons almost equiprobably, provided that the sum of their energies is equal to the energy of the original photon. If, alternatively, $\zeta\ll1$, then, in general, this is not true. However, let us turn our attention to the approximation of the total splitting probability~\eqref{photonSplittingTotalProbabilityApproximation}. The photon splitting probability increases with~$\zeta$; at small~$\zeta$, the probability~$W(k,\zeta)\propto\zeta^6$. It thus follows that the splitting will take place mainly near the pair generation threshold~$\zeta=1$. At small~$\zeta$, the total splitting probability is so low that the specific form of the differential probability~$w(k_1,k_2)$ is unimportant, provided that its integration according to~\eqref{photonSplittingTotalProbability} yields an exact value for the total probability~$W(k)$. Therefore, we will assume below that the photon decays equiprobably into two lower-energy photons even at small~$\zeta$.

If the $\perp$-polarized photons split equiprobably into any two $\parallel$-polarized photons, then
    \begin{equation}
    \label{equiprobableSplitting}
        w(k_1,k_2)=w(k_1+k_2).
    \end{equation}
Given Eq.~\eqref{equiprobableSplitting}, the system of equations~\eqref{photonSplittingSystem} transforms to
    \begin{equation}
    \label{equiprobablePhotonSplittingSystem}
        \begin{aligned}
            &\dot{f}_\perp(k)\!\!\!\!&=&-\frac{k}{2}w(k)f_\perp(k),\\
            &\dot{f}_\parallel(k)\!\!\!\!&=&\int\limits_k^\infty w(k')f_\perp(k')\,dk'.
        \end{aligned}
    \end{equation}
Expressing the product~$w(k)f_\perp(k)$ from the first equation in~\eqref{equiprobablePhotonSplittingSystem} and substituting it into the second equation yields
    \begin{equation}
    \label{ParallelPhotonDistributionEquation}
        \frac{\partial f_\parallel(k,\zeta)}{\partial\zeta}=
        -2\frac{\partial}{\partial\zeta}\int\limits_k^\infty f_\perp(k',\zeta)\frac{dk'}{k'}.
    \end{equation}
Here, we used the fact that $\partial/\partial t=\partial/\partial l=[k/2\rho_0]\,\partial/\partial\zeta$. We are interested in the value
of the distribution function for $\parallel$-polarized photons at~$\zeta=1$. Below, we will see that the generation of electron–positron pairs by curvature photons in the strong magnetic field of a magnetar actually takes place at~$\zeta=1$. It should be noted that the splitting at~$\zeta=1$ has already finished and there are virtually no $\perp$-polarized photons (see Eqs.~\eqref{thresholdValueOfPerpendicularPhotonDistribution}~and~\eqref{thresholdValueOfKForPhotonSplitting}). Therefore, we should set~$f_\perp(k,1)=0$. Integrating~\eqref{ParallelPhotonDistributionEquation} by taking this circumstance into account, we find that
    \begin{equation}
    \label{ParallelPhotonDistribution}
        f_\parallel(k,1)\approx f_\parallel(k,\infty)=f_\parallel(k,0)+2\int\limits_k^\infty f_\perp(k',0)\frac{dk'}{k'}.
    \end{equation}
We see from Eq.~\eqref{ParallelPhotonDistribution} that for equiprobable splitting of $\perp$-polarized photons (see~\eqref{equiprobableSplitting}), the asymptotic expression for the distribution function of $\parallel$-polarized photons does not depend on the specific form of~$w(k)$.
\section*{THE ELECTRON–POSITRON PLASMA GENERATION}
\addcontentsline{toc}{section}{The Electron-Positron Plasma Generation}

Let us derive the specific form of the operator~$Q$ that describes the generation of relativistic electrons and positrons by high-energy photons for a weak magnetic field, $B\ll1$, characteristic of ordinary pulsars,
and for a strong magnetic field, $B\gtrsim1$, characteristic of magnetars.

Consider the case of~$B\ll1$. If we pass from the variable~$\chi'$ to~$\mu'=2/k\chi'$ and from the variable~$k$ to~$\mu=2/k\chi$ in Eq.~\eqref{Q} for the operator~$Q$, then it will transform to
    \begin{equation}
    \label{Q1}
        Q=2\alpha b B\rho_0\int d\chi\,\delta\Big(\gamma-\frac{1}{\chi}\Big)\,\int\limits_0^1\frac{d\mu}{\mu^2}\,q_0\Bigl(\frac{2}{\mu\chi}\Bigr)\exp\Bigl(-a\mu-\alpha b\rho_0\chi^2\mu^2
        \int\limits_\mu^1\frac{d\mu'}{\mu'^3}\,B'e^{-a'\mu'}\Bigr),
    \end{equation}
where by the coefficient~$b$ we will mean either~$b_\parallel$ or~$b_\perp$ (see~\eqref{weakFieldPhotonAttenuationCoefficient}). When criterion~\eqref{chiCriterion} is met, we may assume that $B'$ does not depend on the variable~$\mu'$ and coincides in value with the quantity~$B$. Integrating the expression in the exponent by parts and retaining the principal term (which is legitimate, because, as we will see below, the integrand differs noticeably from zero only at~$\mu\ll1$), we obtain
    \begin{equation}
    \label{Q2}
        Q=2a p\int\limits_0^a\frac{d y}{y^2}\,q_0\Bigl(2\gamma\frac{a}{y}\Bigr)\exp\Bigl(-y-p\frac{e^{-y}}{y}\Bigr).
    \end{equation}
Here, we use the notation
    \begin{equation}
    \label{a}
        p=\frac{\alpha b\rho_0B}{\gamma^2},\qquad a=\frac{4}{3B}.
    \end{equation}

Let us determine the maximum of the exponent in Eq.~\eqref{Q2}. Assuming that this maximum is reached at~$y=\Lambda\gg1$, we find that~$\Lambda$ must be a solution of the equation $\Lambda=\Lambda_0-\ln\Lambda$, where~$\Lambda_0=\ln p$. At large~$\Lambda$, the solution of the equation is given with a sufficient accuracy by
    \begin{equation}
    \label{lambda}
        \Lambda\approx\Lambda_0-\ln\Lambda_0.
    \end{equation}

For typical radius of curvature of the field lines $\rho_0\gtrsim10^{19}$, magnetic field $B\simeq0.01{-}0.1$, and energy of the produced particles $\gamma\lesssim10^4$, $\Lambda\simeq15{-}20$. Since the coefficient~$b$ appears under the logarithm, it makes no sense to distinguish~$\Lambda_\parallel$ and~$\Lambda_\perp$ for different directions of photon polarization. Below, we will assume that $b=(b_\parallel+b_\perp)/2\approx 0.23$.

Expanding the exponent at~$y=\Lambda$ gives~$\exp(-y-pe^{-y}/y)\approx\exp(-\Lambda-(y-\Lambda)^2/2)$ at~$\Lambda\gg1$. We see that this function has a sharp peak at~$y=\Lambda$ and that its variance~$\sigma^2=1$. Therefore, the peak width is equal to unity and does not depend on~$\Lambda$.

As we see from~\eqref{Q2}, if $a<\Lambda$, then~$Q=0$. If,
alternatively, $a>\Lambda$, then for a weak magnetic field,
    \begin{equation}
    \label{Q3}
        Q=\frac{2a}{\Lambda}q_0\Bigl(2\gamma\frac{a}{\Lambda}\Bigr),\qquad B\ll1.
    \end{equation}
We used the fact that at~$\Lambda\gg1$, the integral
    \begin{equation*}
    \label{I}
        \int\limits_0^\infty\frac{dy}{y}\,p\frac{e^{-y}}{y}\exp\Bigl(-p\frac{e^{-y}}{y}\Bigr)\approx\frac{1}{\Lambda},
    \end{equation*}
where~$\Lambda$ is defined by~\eqref{lambda}.

Let us now consider the case of~$B\gtrsim1$. Using~\eqref{Q} and~\eqref{strongFieldParallelPolarizationPhotonAttenuationCoefficient}, we can easily obtain
    \begin{equation*}
    \label{Q4}
    \begin{split}
        Q&=\frac{\alpha\rho_0B}{\gamma^2}\int\limits_0^{\pi/2}d\eta\,\cos{\eta}\exp\Bigl(-\frac{2}{B\cos^2\eta}\Bigr)
        q_0\Bigl(\frac{2\gamma}{\cos\eta}\Bigr)\\
        &\times\exp\left(-\frac{\alpha\rho_0B}{2\gamma^2}\cos^2\eta\int\limits_0^\eta\exp\Bigl(-\frac{2}{B\cos^2\eta'}\Bigr)\,d\eta'\right),
    \end{split}
    \end{equation*}
where $\cos\eta=2/k\chi$ and the angle~$\chi$ is related to~$\gamma$ by Eq.~\eqref{gammaOnChiDependence}. Since the condition $\rho_0/\gamma^2\gg1$ is always satisfied, the integrand differs noticeably from zero only at the lower limit of integration. Expanding the integrand at~$\eta=0$ and performing integration from zero to infinity, we find that for a strong magnetic field,
    \begin{equation}
    \label{Q5}
        Q=2q_0(2\gamma),\qquad B\gtrsim1.
    \end{equation}

The physical meaning of Eqs.~\eqref{Q3} and~\eqref{Q5} is simple. At~$B\ll1$, an emitted photon with energy~$k$ will traverse the distance~$l_f=l_t\,a/\Lambda$, where~$l_t=2\rho_0/k$, and will then immediately generate an electron–positron pair at this point. Thus, $l_f$ is the photon mean free path in a magnetic field. If~$B\gtrsim1$, then~$l_f=l_t$, i.e., for the strong magnetic fields of magnetars, the generation of pairs by photons takes place immediately after the threshold has been reached.

When condition~\eqref{chiCriterion} is satisfied, the value of~$q_0(k)$ at the point of photon emission is equal to the value of~$q(k)$ at the point of photon absorption, i.e., $q_0(k)\approx q(k)$. Let us derive an expression for the function~$q(k)$. For a weak magnetic field,
    \begin{equation}
    \label{weakFieldSmallQ}
        q(k)=q^{(c)}(k)+q^{(s)}(k),
    \end{equation}
where the functions $q^{(c)}(k)$ and~$q^{(s)}(k)$ give the numbers
of curvature photons and synchrophotons emitted by a unit volume per unit time. We will distinguish the functions $q_+^{(c)}(k)$ for $\parallel$-polarized photons and $q_-^{(c)}(k)$ for $\perp$-polarized photons:
    \begin{equation}
    \label{weakFieldSmallQForCurvaturePhotons}
        q_\pm^{(c)}(k)=\int\limits_0^\infty P_\pm^{(c)}(\gamma,k)F_0(\gamma)\,d\gamma.
    \end{equation}
The lower limit in Eq.~\eqref{weakFieldSmallQForCurvaturePhotons} may be assumed to be zero, because criterion~\eqref{classicalSyncrotronTheoryApplicabilityCriterion} is met.

To calculate the specific form of the function~$q^{(c)}(k)$ for~$B\ll1$ and~$B\gtrsim1$, we must know the form of the distribution function for the primary particles~$F_0(\gamma)$ after their acceleration in the inner gap. When the boundary condition~\eqref{boundaryCondition} is satisfied, the primary particle distribution function is
    \begin{equation}
    \label{primaryParticlesDistributionFunction}
        F_0(\gamma)=N_0\frac{B}{B_0}\delta(\gamma-\gamma_0),
    \end{equation}
where $B$ is the magnetic field strength at the point under consideration, $B_0$ and~$N_0$ are the magnetic field strength and the plasma density at the stellar surface. We will assume that~$N_0$ is proportional to the
Goldreich–Julian density $|N_0^{GJ}|=\Omega B_0|\cos\theta|/2\pi ce$ at the stellar surface, where $\theta$ is the inclination of the magnetic dipole axis to the neutron star rotation axis, with the proportionality factor being determined by the total longitudinal current~$j_{\parallel}$ flowing in the
magnetosphere, so that $N_0=i_0|N_0^{GJ}|$, where $i_0=2\pi j_{\parallel}/\Omega B_0|\cos\theta|\leqslant1$.

For the weak magnetic fields of ordinary pulsars,
    \begin{equation}
    \label{q1}
        q^{(c)}(k)=q_+^{(c)}(k)+q_-^{(c)}(k)=\frac{N_0}{\sqrt{3}\,\pi}\frac{B}{B_0}\frac{\alpha}{\gamma_0^2}\varphi\Bigl(\frac{2k\rho}{3\gamma_0^3}\Bigr),\qquad B\ll1,
    \end{equation}
where the function $\varphi(x)$ is defined as
    \begin{equation}
    \label{phi}
        \varphi(x)=\frac{\varphi_+(x)+\varphi_-(x)}{2}=\int\limits_x^\infty K_{5/3}(y)\,dy
    \end{equation}
and has the asymptotics
    \begin{equation}
    \label{phiAsymptotics}
        \varphi(x)=
        \begin{cases}
            2^{2/3}\Gamma\!\!\left(\!\displaystyle\frac{2}{3}\!\right)x^{-2/3},\quad x\rightarrow0\\            \sqrt{\displaystyle\frac{\pi}{2x}}\,e^{-x},\quad x\rightarrow\infty.
        \end{cases}
    \end{equation}
If, alternatively, $B\gtrsim1$, then the photon decay processes are important. Therefore, Eq.~\eqref{ParallelPhotonDistribution} must be used to find the form of the function~$q^{(c)}(k)$. Then,
    \begin{equation}
    \label{q2}
        q^{(c)}(k)=\frac{N_0}{2\sqrt{3}\,\pi}\frac{B}{B_0}\frac{\alpha}{\gamma_0^2}g\Bigl(\frac{2k\rho}{3\gamma_0^3}\Bigr),\qquad B\gtrsim1,
    \end{equation}
where the function~$g(x)$ is defined as
    \begin{equation}
    \label{g}
        g(x)=\varphi_+(x)+2\int\limits_x^\infty\varphi_-(y)\frac{dy}{y}.
    \end{equation}
Its asymptotics are
    \begin{equation}
    \label{gAsymptotics}
        g(x)=
        \begin{cases}
            2^{2/3}3\,\Gamma\!\!\left(\!\displaystyle\frac{2}{3}\!\right)x^{-2/3},\quad x\rightarrow0\\            \sqrt{\displaystyle\frac{2\pi}{x}}\,e^{-x},\quad x\rightarrow\infty.
        \end{cases}
    \end{equation}

For $B\gg1$, no synchrophotons are emitted, since the particles are produced straight at the zeroth Landau level. Therefore, we should set $q^{(s)}(k)=0$ in Eq.~\eqref{weakFieldSmallQ}. We see that only the first generation of particles produced by curvature photons exists in the magnetosphere of a magnetar. If, alternatively, $B\ll1$, as in the magnetosphere of a pulsar, then the contribution from synchrophotons is significant and $q^{(s)}(k)$ is nonzero; therefore, the second generation of particles can be produced by synchrophotons. For comparison with the magnetosphere of a magnetar, we will need only the first-generation particle distribution function.

Given the conditions~$\gamma\gg1$ and~$\textrm{div\,}{\bf B}=0$, the kinetic equation~\eqref{KineticEquation} for the steady-state case transforms to
    \begin{equation}
    \label{KineticEquation2}
        {\bf b}\nabla F'_1=Q',
    \end{equation}
where $F'_1=F_1/B$ and~$Q'=Q/B$ are, respectively, the distribution function for the first generation of particles and the particle source divided by the magnetic field strength. If~$B\ll1$, then~$Q$ can be determined from~\eqref{Q3} and~\eqref{q1}. So, Eq.~\eqref{KineticEquation2} takes the form
    \begin{equation}
    \label{weakFieldKineticEquation}
        \left(\frac{\partial}{\partial z}+\frac{3r}{2z}\frac{\partial}{\partial r}\right){F'_1}^{(w)}=\frac{2}{\sqrt{3}\,\pi}\frac{\alpha}{\gamma_0^2}\frac{N_0}{B_0}\frac{a}{\Lambda}\,\varphi\Bigl(\frac{4\gamma\rho}{3\gamma_0^3}\frac{a}{\Lambda}\Bigr),\qquad B\ll1.
    \end{equation}

We see from Eq.~\eqref{phiAsymptotics} for the asymptotics of~$\varphi(x)$ that the first-generation particle distribution function $F'_1(\gamma)$ for a weak magnetic field differs noticeably from zero only at~$\gamma<\gamma_{\max}^{(w)}=3\gamma_0^3\Lambda/4a\rho$. At~$\gamma>\gamma_{\max}$, the function~$F'_1(\gamma)$ decreases exponentially and gives no contribution to the total plasma density. We will be interested in the range~$\gamma<\gamma_{\max}^{(w)}$. In this case, the solution to Eq.~\eqref{weakFieldKineticEquation} is
    \begin{equation}
    \label{weakFieldSolution}
        {F'_1}^{(w)}(\gamma)=\frac{3^{11/6}}{10\pi}\,\Gamma\!\!\left(\!\displaystyle\frac{2}{3}\!\right)
        \alpha\frac{N_0}{B_0}\left(\frac{a_0}{\Lambda}\right)^{1/3}R^{-1/3}R_s^{2/3}\gamma^{-2/3}\left[\left(\frac{z}{R_s}\right)^{5/3}-1\right],
    \end{equation}
    \begin{equation*}
        B\ll1,\qquad\gamma<\gamma_{\max}^{(w)}=3\gamma_0^3\Lambda/4a\rho.
    \end{equation*}
Here, $R_s$ is the neutron star radius, $a_0=4/3B_0$, $B_0$ is the surface magnetic field, and~$R=z^3/r^2$ is the integral of motion that is conserved along a magnetic field line. We set the boundary condition~$F'_1(\gamma)=0$ at~$z=R_s$, because the curvature photons producing the first generation of particles just begin to appear near the stellar surface.

Let us now consider the case of a strong magnetic field, $B\gtrsim1$. Using Eqs.~\eqref{Q5} and~\eqref{q2}, we obtain
    \begin{equation}
    \label{strongFieldKineticEquation}
        \left(\frac{\partial}{\partial z}+\frac{3r}{2z}\frac{\partial}{\partial r}\right){F'_1}^{(s)}=\frac{1}{\sqrt{3}\,\pi}\frac{\alpha}{\gamma_0^2}\frac{N_0}{B_0}\,g\Bigl(\frac{4\gamma\rho}{3\gamma_0^3}\Bigr),\qquad B\gtrsim1.
    \end{equation}

Just as in the previous case, we will be interested in the form of the distribution function~$F'_1(\gamma)$ only at~$\gamma<\gamma_{\max}^{(s)}=3\gamma_0^3/4\rho$ (see asymptotics~\eqref{gAsymptotics}). Under this condition, Eq.~\eqref{strongFieldKineticEquation} has the following solution:
    \begin{equation}
    \label{strongFieldSolution}
        {F'_1}^{(s)}(\gamma)=\frac{3^{17/6}}{8\pi}\,\Gamma\!\!\left(\!\displaystyle\frac{2}{3}\!\right)
        \alpha\frac{N_0}{B_0} R^{-1/3}R_s^{2/3}\gamma^{-2/3}\left[\left(\frac{z}{R_s}\right)^{2/3}-1\right],
    \end{equation}
    \begin{equation*}
        B\gtrsim1,\qquad\gamma<\gamma_{\max}^{(s)}=3\gamma_0^3/4\rho.
    \end{equation*}

We see that the first-generation particle distribution functions for both weak and strong magnetic fields are $F'_1(\gamma)\propto\gamma^{-2/3}$ and have the same order of magnitude. Near the surface, the functions increase linearly in~$(z-R_s)$, with their ratio being
    \begin{equation*}
    \label{DistributionRatio}
        \frac{{F'_1}^{(w)}(\gamma)}{{F'_1}^{(s)}(\gamma)}=\frac{2}{3}\left(\frac{a_0}{\Lambda}\right)^{1/3}\simeq1.
    \end{equation*}
The density ratio will be slightly different, because~$\gamma_{\max}$ depends on the magnetic field strength, and is smaller in a weak field than in a strong one. Integrating Eqs.~\eqref{weakFieldSolution} and~\eqref{strongFieldSolution} over the particle Lorentz factor from~$\gamma_{\min}$ to~$\gamma_{\max}$ yields
    \begin{equation}
    \label{weakFieldPlasmaDensity}
        {N'_1}^{(w)}=\frac{3^{17/6}}{10\pi}\,\Gamma\!\!\left(\!\displaystyle\frac{2}{3}\!\right)
        \alpha\frac{N_0}{B_0}\left(\frac{a_0}{\Lambda}\right)^{1/3}R^{-1/3}R_s^{2/3}{\gamma_{\max}^{(w)}}^{1/3}\Biggl[1-\left(\frac{\gamma_{\min}}{\gamma_{\max}^{(w)}}\right)^{1/3}\Biggr]\left[\left(\frac{z}{R_s}\right)^{5/3}-1\right],
    \end{equation}
    \begin{equation*}
        B\ll1,\qquad\gamma_{\min}<\gamma<\gamma_{\max}^{(w)};
    \end{equation*}
    \begin{equation}
    \label{strongFieldPlasmaDensity}
        {N'_1}^{(s)}=\frac{3^{23/6}}{8\pi}\,\Gamma\!\!\left(\!\displaystyle\frac{2}{3}\!\right)
        \alpha\frac{N_0}{B_0} R^{-1/3}R_s^{2/3}{\gamma_{\max}^{(s)}}^{1/3}\Biggl[1-\left(\frac{\gamma_{\min}}{\gamma_{\max}^{(s)}}\right)^{1/3}\Biggr]\left[\left(\frac{z}{R_s}\right)^{2/3}-1\right],
    \end{equation}
    \begin{equation*}
        B\gtrsim1,\qquad\gamma_{\min}<\gamma<\gamma_{\max}^{(s)}.
    \end{equation*}

The minimum particle Lorentz factor $\gamma_{\min}$ is determined
by the maximum angle~$\chi_{\max}$ (see~\eqref{chiCriterion}) and is
equal to $\gamma_{\min}=1/\chi_{\max}\simeq100$. Since the true values
of the distribution function~$F(\gamma)=BF'(\gamma)$ and the
plasma density~$N=BN'$, the ratios of the distribution functions and densities near the surface for strong and weak magnetic fields are
    \begin{equation}
    \label{DistributionRatio2}
        \frac{F_1^{(w)}(\gamma)}{F_1^{(s)}(\gamma)}=\frac{2}{3}\left(\frac{a_0}{\Lambda}\right)^{1/3}\frac{B_0^{(w)}}{B_0^{(s)}},
    \end{equation}
    \begin{equation}
    \label{DensityRatio}
        \frac{N_1^{(w)}}{N_1^{(s)}}=\frac{2}{3}\frac{B_0^{(w)}}{B_0^{(s)}},
    \end{equation}
where $a_0=4/3B_0^{(w)}$, while~$B_0^{(w)}$ and~$B_0^{(s)}$ are the
surface magnetic field strengths for an ordinary pulsar and a magnetar, respectively. In calculating ratios~\eqref{DistributionRatio2} and~\eqref{DensityRatio}, we used the conditions~$\gamma_{\min}\ll\gamma_{\max}$ and $(z-R_s)/R_s\ll1$.

A factor that increases when~$z\rightarrow\infty$ appears in Eqs.~\eqref{weakFieldSolution}, \eqref{strongFieldSolution}, \eqref{weakFieldPlasmaDensity}, and~\eqref{strongFieldPlasmaDensity}. This is because we assumed condition~\eqref{chiCriterion} to be satisfied, i.e., we
assumed that the mean free path of a curvature photon is small; therefore, the distribution functions and magnetic field strengths at the points of photon emission and absorption coincide. However, if~$\chi\sim\chi_{\max}=z_0/\rho_0$, then the mean free path is large and the electron–positron pair generation takes place at distances much larger than~$z_0$. Therefore, the rate of particle generation decreases with increasing~$z$ and the distribution function~$F'_1(\gamma)$ reaches a stationary value when a certain characteristic value of~$z_c$ is reached.

It is easy to estimate the characteristic value of~$z_c$ from energy considerations. The energy density of the first generation of particles in the magnetosphere of a magnetar cannot exceed the energy density of the primary particles accelerated in the inner gap. The energy density normalized to the magnetic field strength is
    \begin{equation}
    \label{energyDensity}
        {E'_1}^{(s)}=\frac{3^{23/6}}{32\pi}\,\Gamma\!\!\left(\!\displaystyle\frac{2}{3}\!\right)
        \alpha\frac{N_0}{B_0} R^{-1/3}R_s^{2/3}{\gamma_{\max}^{(s)}}^{4/3}\Biggl[1-\left(\frac{\gamma_{\min}}{\gamma_{\max}^{(s)}}\right)^{4/3}\Biggr]\left[\left(\frac{z}{R_s}\right)^{2/3}-1\right].
    \end{equation}
Given Eq.~\eqref{energyDensity}, the condition~${E'_1}^{(s)}<N_0\gamma_0/2B_0$ transforms to
    \begin{equation}
    \label{energyCriterion}
        \frac{\gamma_{\max}^{(s)}}{\gamma_{\min}}\left[\left(\frac{z}{R_s}\right)^{2/3}-1\right]<\frac{1}{2\alpha},\qquad\gamma_{\min}=\frac{\rho}{R_s}\simeq100.
    \end{equation}
For a typical primary particle energy~$\gamma_0\simeq10^7{-}10^8$ the maximum energy of the first-generation particles is~$\gamma_{\max}^{(s)}\simeq10^3{-}10^4$. Therefore, the generation of particles terminates at heights~$z_c-R_s\simeq10R_s$.

An upper limit for the plasma multiplicity~$\lambda=N_1^{\pm}/N_{GJ}$ in the magnetosphere of a magnetar can be easily obtained from Eq.~\eqref{energyCriterion}:
    \begin{equation*}
    \label{maxMultiplicity}
        \lambda_{\max}^{(s)}=2i_0\frac{\gamma_0}{\gamma_{\max}^{(s)}}\lesssim10^3-10^4.
    \end{equation*}

However, it should be noted that the electron–positron pair generation will be effective only if the inclination of the photon wave vector to the magnetic field direction is small compared to the maximum possible angle~$\chi_{\max}$ (see~\eqref{chiCriterion}). This is equivalent to the fact that the photon mean free path is much smaller than the distance from the dipole center to the point of photon emission, i.e., $l\ll z_0$. If, however, $l$ becomes equal to $z_0$, then according to the factors mentioned above, the generation efficiency of a secondary plasma will be low. It is easy to find a criterion under the satisfaction of which the first generation of particles will be produced. For this purpose, we must equate the minimum possible mean free path~$l_{f}=\rho/\gamma_{\max}^{(c)}$ to the distance from the stellar center~$z_0=R_s$. In this case, we must take into account the fact that the primary particle energy~$\gamma_0$ appearing in the expression for~$\gamma_{\max}^{(c)}$ is determined by the electric potential in the gap $\Psi$. The potential distribution near the stellar surface can be obtained by solving the Poisson equation
    \begin{equation*}
    \label{PoissonEquation}
        \left(\frac{1}{r}\frac{\partial}{\partial r}r\frac{\partial}{\partial r}+\frac{\partial^2}{\partial z^2}\right)\Psi=-4\pi(\rho_e-\rho_{GJ}),
    \end{equation*}
where $\rho_e$ is the electric charge density and $\rho_{GJ}=-{\bf\Omega}{\bf B}/2\pi c$ is the Goldreich–Julian charge density. The boundary conditions are a zero potential at the stellar surface~$z=R_s$ and at the surface defined by the last closed field lines, $r=R_p$, where~$R_p=R_s^{3/2}/R_c^{1/2}\simeq 100$~m is the polar cap radius. The solution is given in the form of a series in Bessel functions \citep{BeskinEtal1993}. If the distance from the stellar surface $(z-R_s)>R_p/\mu_1$, where~$\mu_1\approx2.4$ is the first root of the zeroth-order Bessel function, then the potential ceases to change along~$z$ and the expression for the radial potential distribution in dimensionless units is
    \begin{equation*}
    \label{potentialDistribution}
        \Psi(r)=\cos\theta(1-i_0)\frac{B_0}{2}\frac{R_s^3}{R_c^2}\left[1-\Bigl(\frac{r}{R_p}\Bigr)^2\right].
    \end{equation*}

Using the expression for~$\gamma_{\max}^{(s)}$ and given that~$\gamma_0=|\Psi(r)|$, it is easy to obtain the radial distribution of the mean free path for the photons emitted by primary particles near the stellar surface:
    \begin{equation}
    \label{MFPDependenceOnR}
        l_f(r)=\left(\frac{8}{3B_0}\frac{R_c^2}{R_s^3}\right)^3\frac{R_sR_c}{\left(|\cos\theta|(1-i_0)\right)^{3}}
        \Bigl(\frac{r}{R_p}\Bigr)^{-2}\left[1-\Bigl(\frac{r}{R_p}\Bigr)^2\right]^{-3}.
    \end{equation}

We see from Eq.~\eqref{MFPDependenceOnR} that the mean free path $l$ reaches a minimum at~$r=R_p/2$. Equating $l$ to the stellar radius~$R_s$, we can obtain a lower limit for the surface magnetic field strength of a magnetar at which the generation of secondary particles is still possible:
    \begin{equation}
    \label{bCriterion}
        B_0>\frac{8}{3}\frac{R_c^{7/3}}{R_s^3}\phi,\qquad\qquad B_0\gtrsim1,
    \end{equation}
    \begin{equation*}
        \phi=\frac{2^{8/3}}{3|\cos\theta|(1-i_0)}.
    \end{equation*}
If the longitudinal current and the angle between the rotation axis and the magnetic dipole axis are small, then the coefficient~$\phi\approx2.1$. Once the dimensions have been restored, we obtain
    \begin{equation}
    \label{bDimentionalCriterion}
        B_0\gtrsim\left(\frac{P}{1\text{ s}}\right)^{7/3}\left(\frac{R_s}{10\text{ km}}\right)^{-3}\frac{10^{12}}{|\cos\theta|(1-i_0)}\text{ G}.
    \end{equation}
\section*{DISCUSSION}
\addcontentsline{toc}{section}{Discussion}

We may conclude that for each specific rotation period~$P$, there exists some minimum surface magnetic field strength at which an effective generation of a secondary plasma begins. Qualitatively, this can be understood in the following way. The distribution function of electrons and positrons in the magnetosphere of a magnetar is limited below by~$\gamma_{\min}$ and
above by~$\gamma_{\max}^{(s)}$. As the radius of curvature of the field line increases, the minimum particle Lorentz factor increases, while the maximum one decreases. There exists some critical radius of curvature at which these become equal. Since the radius of curvature increases with neutron star rotation period, high primary particle energies proportional to the surface magnetic field strength are required for stars with long rotation periods. If we equate $\gamma_{\min}$ and $\gamma_{\max}^{(s)}$, while using a simple
estimate of~$\gamma_0\simeq\ B_0R_s^3/2R_c^2$ for the primary particle energy, and take the radius of curvature at the point of intersection of the last closed field line with the stellar surface, then we will exactly obtain estimate~\eqref{bCriterion} for the field strength but with the coefficient~$\phi=1$.

Equation~\eqref{bDimentionalCriterion} allows us to construct the line in
the~$P{-}{\dot P}$~diagram that separates the neutron stars with an active generation of electron–positron pairs in their magnetospheres. If we use the surface magnetic field $B_0\simeq2(P{\dot P}_{-15})^{1/2}\,10^{12}$~G estimated from the formula for the magnetodipole losses and assume the longitudinal current and the angle between the dipole axis and the stellar rotation axis to be small, $j_{\parallel}\ll j_{GJ}=c\rho_{GJ}$ and~$\cos\theta\simeq1$, respectively, then we can easily find that
    \begin{equation}
    \label{PDotP}
        \log{\dot P}_{-15}=\frac{11}{3}\log P-0.54,
    \end{equation}
where $P$ is the neutron star rotation period in seconds
and ${\dot P}_{-15}$ is the period derivative in~$10^{-15}\text{ s s}^{-1}$.

\hyperlink{pdotp}{The figure} presents the~$P{-}{\dot P}$~diagram in which line~\eqref{PDotP} was drawn. When constructing the diagram, we used data from the ATNF catalog \citep{ManchesterEtal2005}. We see that all magnetars are located to the left of this line and fairly close to it. This suggests that secondary particles can be generated in principle in their magnetospheres. Therefore, we have no reason to believe that the generation of radio emission is suppressed in magnetars. Indeed, we showed here that the photon splitting, which was previously considered to be a pair generation suppression factor, is not such a factor. Although the splitting undoubtedly takes place in the superstrong magnetic fields of magnetars, it leads only to a strong polarization of the curvature photons, but not to a decrease in their number or energy. The corresponding change in the maximum characteristic energy of the secondary particles cannot be a suppression factor either, because this energy for magnetars is even higher than that for radio pulsars. The ratios of the first-generation particle densities to the magnetic field strength in the cases of weak and strong fields are equal in order of magnitude. Therefore, the first-generation plasma density in the magnetosphere of a magnetar is even higher than that in the magnetosphere of a pulsar with a weak magnetic field.

The fact that no secondary generation of particles is produced in a strong magnetic field can serve as a factor of suppression but incomplete one. Since the first-generation particles are produced straight at the zeroth Landau level, there is no emission of the synchrophotons that produce the second-generation particles. The contribution from synchrophotons is known to dominate in the magnetospheres of radio pulsars. Numerical simulations of plasma generation cascades \citep{DaughertyHarding1982,HibschmanArons2001} suggest that the total secondary particle distribution function has the form of a power law~$\gamma^{-(1.5-2)}$. In contrast, only the first generation of secondary particles with a distribution function proportional to~$\gamma^{-2/3}$ exists for magnetars. It thus follows that, in general, the plasma generation multiplicity~$\lambda=N/N_{GJ}$ in the magnetosphere of a magnetar is smaller than that in the magnetosphere of an ordinary radio pulsar.

The already mentioned closeness of the minimum and maximum Lorentz factors for the first-generation particles, which can lead to a sharp decrease in the electron–positron plasma density (see Eq.~\eqref{strongFieldPlasmaDensity}), can serve as another plasma generation suppression factor. Indeed, all magnetars are fairly close to line~\eqref{PDotP}. In addition, the plasma generation efficiency depends on the longitudinal current flowing in the magnetosphere and on the angle between the axes. Obviously, this dependence is particularly sensitive to a change in parameters near the threshold line. The threshold magnetic field strength~\eqref{bDimentionalCriterion} also increases with current~$j_{\parallel}$ and angle~$\theta$. Therefore, adjacent magnetars in the~$P{-}{\dot P}$~diagram may or may not exhibit an activity in the radio frequency band. For example, according to \citet{MalofeevEtal2004}, weak radio pulses are observed from AXP~1E~2259+586 at a low frequency of 111~MHz. According to \citet{CamiloEtal2006}, an intense radio emission at high frequency 0.7--42~GHz with a linear polarization close to 100\% \citep{KramerEtal2007,CamiloEtal2007} emerged from AXP~XTE~1810–197 after its X-ray outburst in~2003. The differences in the frequency and intensity of radio emission from these two magnetars could be explained by a difference in plasma density and multiplicity, since the first magnetar is considerably closer to the threshold line than the second one. However, no radio emission from other magnetars has been observed so far, which may be indicative of the influence of some other factors. Nonetheless, undoubtedly, one might expect the possible presence of weak radio pulses from them.

For magnetars close to the threshold line~\eqref{PDotP} in the~$P{-}{\dot P}$~diagram, the plasma density is low compared to that in the magnetospheres of ordinary pulsars. If, as is customary, the radio frequencies~$\omega$ are defined by the relation~$\omega^2\simeq\omega_p^2/\gamma^3$, where $\omega^2_p=4\pi N e^2/m$ is the square of the plasma frequency, then this leads to a shift in the observed radio frequencies toward the lower values, which is observed \citep{MalofeevEtal2005}.

We see that an effective electron–positron plasma generation is possible in the magnetosphere of a magnetar. However, the absence of radio emission from most magnetars may stem from the fact that the presence of primary photons with a sufficiently high energy that must fall into the polar region is required to trigger a cascade pair generation. But its area is inversely proportional to the light-cylinder radius and, hence, is directly proportional to the rotation rate. At the same photon flux density from the external background, the probability of triggering the plasma generation for magnetars with periods $P\simeq10$~s is a factor of~$10{-}100$ lower than that for pulsars with typical periods $P\simeq 0.3$~s. We probably observed such a triggering in AXP~XTE~1810–197.
\section*{ACKNOWLEDGMENTS}
\addcontentsline{toc}{section}{Acknowledgments}

This work was supported by the Russian Foundation for Basic Research (project no.~05-02-17700) and grant no.~NSh-5742.2006.2 from the President of Russia.
\pagebreak

\clearpage
\begin{figure}[t]
\hypertarget{pdotp}{}
\begin{center}
\includegraphics[width=\textwidth]{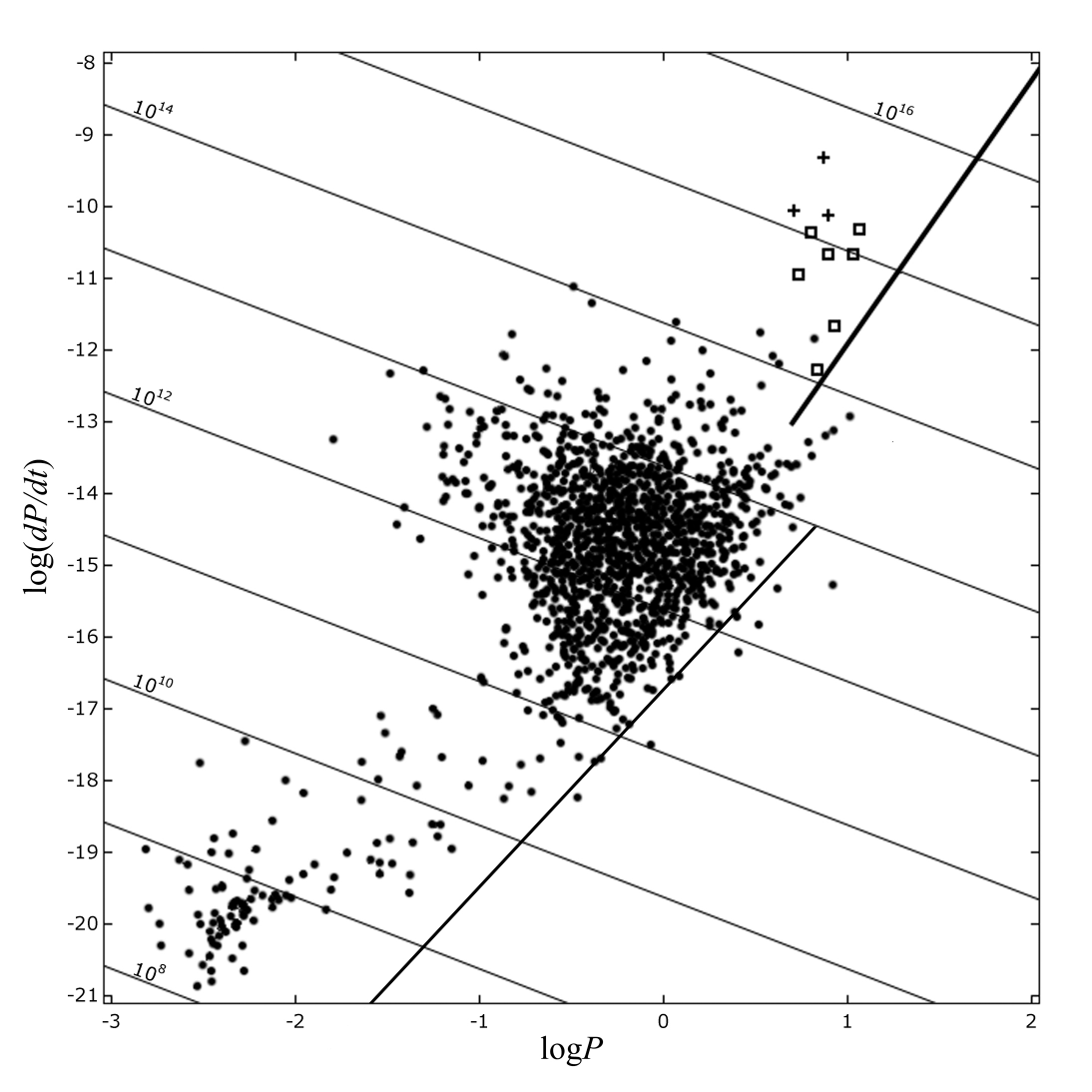}
\caption{Threshold line defining the boundary of effective pair generation in the magnetosphere of a magnetar in the~$P-{\dot P}$~diagram. The crosses, squares, and dots mark SGRs, AXPs, and the remaining pulsars, respectively. For comparison, the death line for ordinary pulsars with a slope of~$11/4$ is also drawn.}
\end{center}
\end{figure}
\end{document}